\documentclass[letterpaper,8pt]{extarticle}  
\usepackage{import}
\usepackage{local}
\usepackage[left=.65in,top=.65in,bottom=.65in,right=.65in]{geometry}
\usepackage{lipsum} 
\usepackage{graphicx}
\usepackage{soul,color}
\usepackage{arydshln}
\usepackage{hyperref}

\title{Mode-selective Raman imaging of metal-organic frameworks reveals surface heterogeneities of single HKUST-1 crystals}

\author{%
\textbf{Matheus Esteves Ferreira \orcidlink{0000-0002-4783-8068},\textcolor{Accent}{\textsuperscript{1,2,*}} %
Mariana Del Grande,\textcolor{Accent}{\textsuperscript{1}} %
Felipe Lopes Oliveira \orcidlink{0000-0002-2231-4567}\textcolor{Accent}{\textsuperscript{4}} 
Rodrigo Neumann Barros Ferreira \orcidlink{0000-0002-4783-8068}\textcolor{Accent}{\textsuperscript{1}} 
Ademir Ferreira da Silva \orcidlink{0009-0006-3832-6411}\textcolor{Accent}{\textsuperscript{1}}
Pamela Costa Carvalho \orcidlink{0000-0002-6801-4762}\textcolor{Accent}{\textsuperscript{1}}
Geisa Lima \orcidlink{0009-0008-0461-4276} \textcolor{Accent}{\textsuperscript{1,4}}
Ado Jorio \orcidlink{0000-0002-5978-2735}\textcolor{Accent}{\textsuperscript{2,3}}
Mathias Steiner \orcidlink{0000-0003-1528-9292}\textcolor{Accent}{\textsuperscript{1}}
}\\
\begin{small}\textcolor{Accent}{\textsuperscript{1}}IBM Research, IBM, Rio de Janeiro, Brazil\\ 
\textcolor{Accent}{\textsuperscript{2}}Graduate Program of Technological
Innovation - PPGIT, Universidade Federal
de Minas Gerais (UFMG), Minas Gerais,
Brazil \\
\textcolor{Accent}{\textsuperscript{3}}Department of Physics, Universidade
Federal de Minas Gerais (UFMG), Minas
Gerais, Brazil\\
\textcolor{Accent}{\textsuperscript{4}}Institute of Chemistry, Universidade
Federal do Rio de Janeiro (UFRJ), Rio de
Janeiro, Brazil\\
\textcolor{Accent}{\textsuperscript{*}}Correspondence: \textcolor{Accent}{m.estevesf@ibm.com} \\ \end{small}
}

\date{}
\begin{document}
\maketitle
\thispagestyle{empty}

\section{Abstract}

\begin{doublespacing}

\noindent
\textbf{\textcolor{Accent}{Metal organic frameworks (MOFs) are nanoporous materials with high surface-to-volume ratio that have potential applications as gas sorbents. Sample quality is, however, often compromised and it is unclear how defects and surface contaminants affect the spectral properties of single MOF crystals. Raman micro-spectroscopy is a powerful tool for characterizing MOFs, yet spatial spectral heterogeneity distributions of single MOF crystals have not been reported so far. In this work, we use Raman micro-spectroscopy to characterize spatially isolated, single crystals of the MOF species HKUST-1. In a first step, we validate HKUST-1's Raman spectrum based on DFT simulations and we identify a previously unreported vibrational feature. In a second step, we acquire diffraction-limited, mode-selective Raman images of a single HKUST-1 crystals that reveal how the spectral variations are distributed across the crystal surface. In a third step, we statistically analyze the measured spectral peak positions and line widths for quantifying the variability occurring within the same crystal as well as between different crystals taken from the same batch. Finally, we explore how multivariate data analysis can aid feature identification in Raman images of single MOF crystals. For enabling validation and reuse, we have made the spectroscopic data and simulation code publicly available.}}

\section{Introduction}

Metal organic frameworks (MOFs), a class of crystalline, nanoporous materials with high surface area and gas adsorption capacity are currently being researched for application in chemical separation \cite{Yaghi2003705, Farha201215016, Furukawa2010424}. By coordinating metallic cluster and organic ligands with a broad range of framework topologies, a large variety of 3D structures can be obtained for supporting specific applications  \cite{Yaghi2003705} \cite{Eddaoudi2001319}. However, the application performance of MOFs is impacted by their chemical quality at molecular scale.

Raman micro-spectroscopy has been used to characterize MOF composition, structure, topology, as well as defects \cite{Gentile202010796, Tittel2023}. In addition, the method was applied to MOFs for investigating molecular interactions \cite{Dhumal20163295, Ethiraj201590, Bae2022, Rivera-Torrente20203614, Hadjiivanov20211286}, framework activation \cite{Prestipino20061337, Kim201510009}, degradation \cite{Todaro2016}, and gas adsorption \cite{Fuchs202314324}. However, MOF Raman studies so far have been performed mainly on bulk samples and have not spatially resolved the spectroscopic signature of single crystals.

For determining how imperfections are distributed in single crystals and if the variations observed at the level of single crystals are representative of the bulk, it is necessary to first spatially isolate individual crystals and then to microscopically resolve the spectroscopic signal at the crystal surface. By applying this approach to a number of crystals taken from the same batch, it is then possible to perform a statistical analysis of their spectroscopic properties. Based on this analysis, the sample heterogeneity can be quantified in terms of spectroscopic metrics, such as spectral shifts and line widths, that are characteristic of single crystals as well as of the bulk.  

In this work, we apply micro-Raman spectroscopy to investigate surface heterogeneity of MOFs in the specific case of HKUST-1. We evaluate its Raman spectrum by comparing measured and simulated data and by identifying the Raman bands that carry information about defects and contaminants. We microscopically resolve the surface of a single crystal and demonstrate how heterogeneity is reflected by mode-selective scattering intensities. We analyze the spectral peak positions and line widths and compare them with the results obtained from 100 spatially isolated single crystals for establishing a Raman spectrum which is representative of the batch. In the following, we briefly introduce the materials and methods used in our investigation.

\section{Materials and Methods}

\subsection*{Sample Preparation}

HKUST-1 metal organic frameworks (copper benzene-1,3,5-tricarboxylate) were purchased as powder (Basolite C300, Sigma-Aldrich). Raman samples were prepared by picking up small amounts of powder using a disposable pipette tip and spreading the material on top of a glass microscope slide. 

\subsection*{Micro-Raman Spectroscopy}

The selection of single HKUST-1 crystals for Raman investigation was done by microscopically mapping an area of 4mm x 4mm on the glass microscope slide, connecting optical brightfield images obtained with a 10x/0.25NA objective (EC-EPIPLAN, Zeiss) in a confocal microscope setup (Alpha 300 RAS, Witec). The resulting sample map was used to identify regions with spatially isolated single crystals. 

Additional images taken with a 63x/0.75NA objective (LD Plan-NEOFLUAR, Zeiss) were used to select crystals with diameters between 1-20$\mu m$ and to measure their respective center positions.

Raman spectra were acquired with a 100x/0.9NA objective (EPIPLAN-Neofluar, Zeiss) in the spectral range of 70-3500 $rel.cm^{-1}$ using a $\lambda$=532nm solid-state laser (Witec) set to 1mW. We used a 600 grooves/mm grating, blazed at 500$nm$ and centered at 2047.09 $rel.cm^{-1}$. The scattered light was detected with a EMCCD camera (Newton, Andor), peltier-cooled to -59 \textdegree C, that has a horizontal resolution of 1600 pixels .

\subsubsection*{Mode-Selective Raman Imaging}

Mode-selective Raman images were taken on spatially isolated, single HKUST-1 crystals with a 100x/0.9NA  microscope objective (EPIPLAN-Neofluar, Zeiss) by raster-scanning a 40 $\mu m$ x40 $\mu m$ sample area with a pixel side length of 200 $nm$. The integration time was set to 0.5s.

\subsection*{Spectral Data Analysis}

A pre-processing step was carried out using the Witec Project 6 software (Oxford Instruments). Firstly, a cosmic ray reduction (CRR) algorithm was executed with filter size and dynamic factor set to 3 and 3, respectively. Secondly, a background removal step was carried out by the shape method, with a size of 110. Then, each spectra $S_i$ was normalized to the intensity of the spectral band at $1006 \pm 25 rel. cm^{-1}$. A non-linear, least-square curve fitting of Lorentzian peak functions was performed using the \textit{LMFIT} python library \cite{newville_2015_11813} for extracting the experimental peak positions, spectral line widths (FWHM), and relative band intensities. The standard deviation was clipped at the 99\% percentile to limit the influence of cosmic rays on the quality of the spectra. 

\subsection*{Principal Component Analysis (PCA)}

Principal Component Analysis (PCA) was performed using the \textit{scikit-learn} python library \cite{scikit-learn}. The PCA parametrization was carried out by defining a 6-dimensional space based on the single-crystal Raman maps. All datasets were transformed using the principal components of the single-crystal data set.

\subsection*{Raman Simulations}

All simulations were based on Density Functional Theory (DFT) under periodic boundary conditions, using the semi-local formulation of the Perdew-Burke-Ernzerhof (PBE) exchange-correlation functional\cite{perdew1996generalized} in combination with the semi-empirical correction for dispersive interactions D3 as proposed by Grimme using the Becke-Johnson damping function variation.\cite{grimme2010consistent, grimme2011effect}

The Kohn-Sham equations were solved using the Quickstep code from the CP2K v2023.1 package\cite{vandevondele2005quickstep, hutter2014cp2k, kuhne2020cp2k} employing the Orbital Transformation (OT)  for wavefunction optimization\cite{vandevondele2003efficient, weber2008direct}. Core electrons were treated with analytical pseudopotentials proposed by Goedecker, Teter, and Hutter (GTH) \cite{goedecker1996separable, hartwigsen1998relativistic}. Valence electrons were expanded in a mixed basis of plane waves and Gaussian waves. The Gaussian basis used was a triple-$\zeta$ with two sets of polarization functions.\cite{vandevondele2007gaussian} For plane waves, a cutoff energy of 1200 Ry was used, mapped into a 5-level grid with a relative cutoff energy of 50 Ry, with the Brillouin zone integration restricted to the $\Gamma$-point.

The atomic positions and cell parameters were derived from the \textit{DOTSOV02\_clean} crystallographic information file (CIF) part of the CoRE MOF 2014 dataset \cite{dalar_nazarian_2020_3986573} . The structure was fully optimized until the total forces were below 1.0 millihartree/bohr (a mean square value below 0.7) and total pressure below 100 bar, using the Broyden–Fletcher–Goldfarb–Shanno (BFGS) minization algorithm with memory limited to 25 steps (L-BFGS). 

The vibrational modes and the intensity of the Raman spectra were both calculated based on the finite differences method under the harmonic approximation, where the potential energy surface was obtained by means of density functional theory using the CP2K v2023.1 \cite{vandevondele2005quickstep, hutter2014cp2k, kuhne2020cp2k} and phonopy \cite{togo2023first, togo2023implementation} packages. More details on the simulations can be found in the supporting Information section, which also includes links to the  data and code repositories.

\section{Results}

\subsection*{Single-crystal Raman spectrum of HKUST-1}

\begin{figure}[b!]
    \centering
    \includegraphics[width=\linewidth]{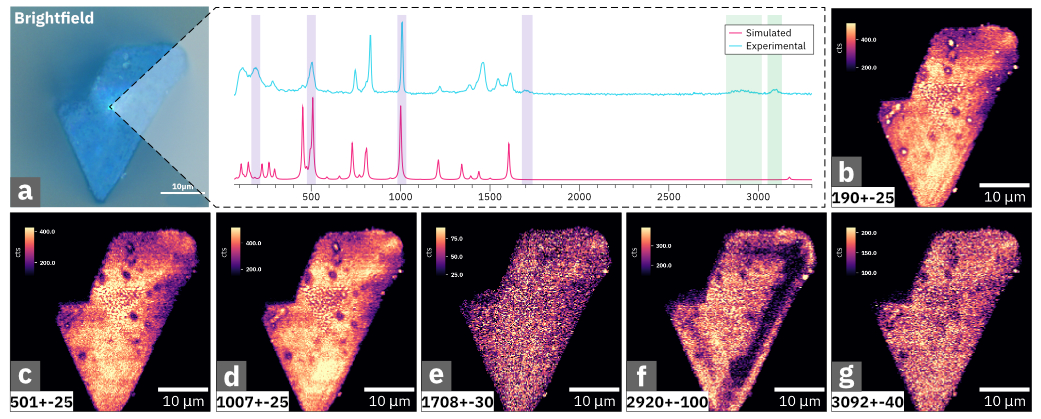}
    \caption{\textbf{Mode-selective Raman imaging of a single, spatially isolated HKUST-1 crystal.} (a) Brightfield microscopy image of a single HKUST-1 crystal. (b-e) Confocal Raman images of the same HKUST-1 crystal representing the integrated intensities of select Raman bands highlighted in the inset. The color-shaded areas (purple: previously reported; green: not reported) highlight the positions of the select Raman bands in Table \ref{tab:1}. (Inset) Averaged Raman spectrum acquired at the center position of a spatially isolated single HKUST-1 crystal [blue] and simulated Raman spectrum [pink] for comparison. }
    \label{fig:raman2d}
\end{figure}

The HKUST-1 sample studied in this investigation consists of octahedral crystals with diameters ranging between 1 and 20 $\mu m$. Figure \ref{fig:raman2d}a shows a brightfield microscopy image of a representative, spatially isolated HKUST-1 crystal. As an experimental reference, we plot in the inset of Figure \ref{fig:raman2d} a HKUST-1 Raman spectrum obtained at the center position of the crystal shown in Figure \ref{fig:raman2d}a . We will now identify the measured HKUST-1 Raman bands and assign the principal vibrational modes based on our simulation results. 

For comparison, we show in the inset of Figure \ref{fig:raman2d} the simulated spectrum which was calculated based on density functional theory, see Methods section. For each mode, we have calculated mode symmetry, center frequency and the respective vector representation along with the scattering intensities for both parallel and perpendicular incidence. Overall, the simulated spectrum in the inset of Figure \ref{fig:raman2d} is composed of 468 independent, vibrational modes and their properties are provided in the supplementary dataset \cite{esteves_ferreira_2024_14165785}.

\begin{figure}[t]
    \centering
    \includegraphics[width=\linewidth]{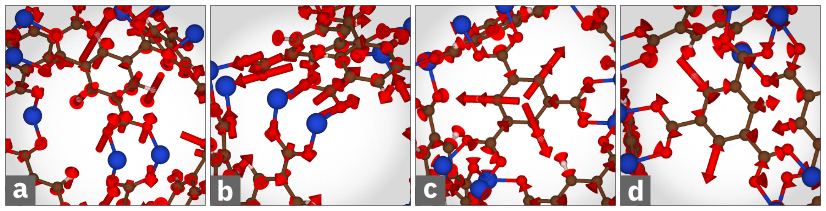}
    \caption{\textbf{Vibrational modes of HKUST-1.} (a-d) The vectors indicate select vibrations summarized in table \ref{tab:1} giving rise to the measured Raman bands at 174-283, 500, 1007, and 3090 $rel. cm^{-1}$, respectively. The respective spectral segments are highlighted in the inset of Figure \ref{fig:raman2d} by purple shadows that are overlaid with the Raman spectra. Copper, oxygen and carbon atoms are represented by blue, red, brown colors, respectively.}
    \label{fig:assig}
\end{figure}

For the purpose of the analysis of surface heterogenities at single-crystal level, we will focus our investigation on a set of Raman modes that carry information about structural defects and impurities. The selected modes are summarized in Table \ref{tab:1} and in Figure \ref{fig:assig} we show their vector representations.  

The modes below 500 $rel. cm^{-1}$ are mainly associated with vibrations related to either $CuO$ or $CuCu$ bonds \cite{Gentile202010796}. In Figure \ref{fig:assig}a, we show the representation of both the bending mode of $OCuO$ group and the torsion of the aromatic ring of the trimesic group. In addition, Figure \ref{fig:assig}b and c visualize the symmetric stretch vibrations of the $CuO$ bonds as well as the in-phase breathing vibration of the aromatic ring. 

Two measured Raman bands centered at 1705 and 2918 $rel. cm^{-1}$, see Figure \ref{fig:raman2d}, do not appear in the simulated spectra. The band at 1705 $rel. cm^{-1}$ was assigned to the out-of-phase bending mode of $COOH$ groups and correlated with local defects of the framework structure originating from carboxyl groups \cite{Gentile202010796}. 

To the best of our knowledge, the spectral features at 2918  and 3090 $rel. cm^{-1}$ have not been previously reported. Based on our simulation results, we have assigned the band at 3090 $rel. cm^{-1}$ to the out-of-phase $CH$ stretch of the aromatic ring in the trimesic group of HKUST-1. The visualization of this vibration is shown in Figure \ref{fig:assig}d.

\begin{table*}[b]
\centering
\begin{tabular}{cccccc}
\hline
\textbf{\begin{tabular}[c]{@{}c@{}}Center\\ (rel. $cm^{-1}$)\end{tabular}} & \textbf{\begin{tabular}[c]{@{}c@{}}$\Delta$ Center\\ (rel. $cm^{-1}$)\end{tabular}} & \textbf{\begin{tabular}[c]{@{}c@{}}Simulated Mode \\ Center \\ (rel. $cm^{-1}$)\end{tabular}} & \textbf{Symmetry} & \textbf{\begin{tabular}[c]{@{}c@{}}Mode\\ Description\end{tabular}} & \textbf{Ref.} \\ \hline
7.96 & 1.15 & 185.07 & $A_{1g}$ & $\tau_{op}(CH)_b$/$\delta(OCuO)$ & \cite{Gentile202010796} \\
500.02 & 1.13 & 508.75 & $A_{1g}$ & $\nu_{ip}(OCuO)$ & \cite{Gentile202010796} \\
1007.00 & 0.93 & 1000.53 & $A_{1g}$ & In phase breathing of the benzene ring (mode 12) & \cite{Gentile202010796} \\
1705.17 & 8.84 & - & - & $\nu_{op}(COOH)$ & \cite{Gentile202010796} \\ \hdashline
2918.20 & 7.53 & - & - & - & This Work \\
3090.84 & 3.55 & 3175.11 & $A_{1g}$ & $\nu_{op}(CH)_b$ & This Work \\ \hline
\end{tabular}%
\caption{\textbf{Vibrational mode assignment of HKUST-1.} Mode assignment of select Raman bands. The horizontal line represents which of the vibrations have been previously unreported. The respective vibrations are visualized in Figure \ref{fig:raman2d}. $\nu$=stretching; $\delta$=bending; $w$=wagging; $ip$=in phase; $op$=Out of phase; $\tau$ = twisting; b=trimesic group.}
\label{tab:1}
\end{table*}

By comparing with our simulation results, we conclude that the band at 2918 $rel. cm^{-1}$ cannot be linked to the HKUST-1 crystal itself. We believe that the feature rather attests to presence of surface contaminants. However, confirming the chemical identity of the adsorbant species would requires further research which is beyond the scope of this paper. 

\subsection*{Mode-selective Raman Imaging of single HKUST-1 crystals}

Having established the Raman spectrum of HKUST-1 at the single-crystal level, we will now investigate if surface heterogeneities emerge at the surface of a single HKUST-1 crystal. By raster scanning the surface of a spatially isolated HKUST-1 crystal with a confocal laser microscope, we have acquired a series of Raman images, see Figures \ref{fig:raman2d} (b-g), for select spectral bands that correspond to the modes summarized in Table \ref{tab:1}. As the Raman signal contains information about specific structural defects and impurities, the images reveal mode-selective heterogeneity at the surface of the same crystal.  

Specifically, in Figure \ref{fig:raman2d}b, which maps the scattering intensities related to the $\delta_{op}(OCuO)$ vibration spectrally centered at 190 $rel. cm^{-1}$, we observe localized centers with enhanced scattering intensities. The observed enhancement of Raman scattering intensity could be related to the presence of metal defects localized close to the crystal surface \cite{portillo2024benefits, müller2017defects} that cause local resonance enhancements \cite{yuan2024metal} or heating effects.

In contrast, Figures \ref{fig:raman2d}c and d, that correspond to bands centered at 501 and 1007 $rel. cm^{-1}$, show reduced scattering intensities at the same locations, leading to contrast inversion with dark centers at the location of the defects. However, all three images Figure \ref{fig:raman2d}b-d show the same, spatially extended heterogeneity at the bottom vertex of the crystal that could be due to changes in surface scattering density. In Figure \ref{fig:sup1} we provide the normalized spatial maps for all bands of HKUST-1.

The image in Figure \ref{fig:raman2d}e, spectrally located at  1708 $rel. cm^{-1}$, shows a rather homogeneous spatial intensity distribtution with low scattering intensities, attesting to the absence of of carboxyl group related defects at the crystal surface\cite{Gentile202010796}.

However, the image in Figure \ref{fig:raman2d}f for the Raman band spectrally centered at 2920 $rel. cm^{-1}$ shows a large, connected area at the right hand side of the crystal surface where the scattering intensity is significantly reduced. Following the spectral assignment in the previous sub-section, we conclude that this signal attests to the presence of contaminants present at the surface of the crystal.

We note that the image for the newly assigned spectral band at 3092 $rel. cm^{-1}$, which is due to the out-of-phase C-H stretch of the benzene ring of the trimesic group, $\nu_{op}(CH)$, is mostly homogeneous throughout the surface, with the exception of some localized defects that are consistent with the other images.

\subsection*{Spectral Variability of Single-Crystal HKUST-1}

Based on the experimental findings that single crystals of HKUST-1 show spatial variations of Raman scattering intensities across their surface, see Figure \ref{fig:raman2d}, the question is how the occurring spectral variations can be quantified. Potentially useful metrics for variability analysis are the spectral peak positions and line widths, respectively, of the measured Raman bands. The experimental values can ultimately serve as a reference for spectroscopic sample-to-sample comparisons.

For analyzing spectral variations within the same crystal, we have measured the Raman spectra of a spatially isolated, single HKUST-1 crystal at 100 different positions equally distributed across the crystal surface. We have statistically evaluated the average spectral peak positions as well as the spectral line widths along with their respective standard deviations. The data were obtained by fitting lorentzian line shape functions to the experimental data, see Methods section.

\begin{figure}[b!]
    \centering
    \includegraphics[width=0.8\linewidth]{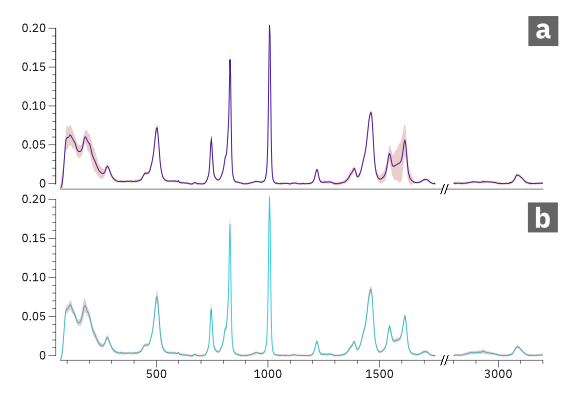}
    \caption{\textbf{Spectral variability of single HKUST-1 crystals.} (a) Raman spectrum obtained by averaging 100 spectra obtained at different positions of the same single crystal. (b) Raman spectrum obtained by averaging 100 spectra recorded at the center positions of 100 different, single crystals of the same batch. Shaded  areas: standard deviation.}
    \label{fig:intra}
\end{figure}

In Figure \ref{fig:intra}a, we plot the spectrum in solid colors and the standard deviation in shaded colors. In the low-frequency regime below 400 $rel. cm^{-1}$, as well as in the spectral ranges between 1100-1800 $rel. cm^{-1}$ and 2800-3100 $rel. cm^{-1}$, we observe high spectral variability. Below 400 $rel. cm^{-1}$ most modes are related to Cu(ii) open metal sites and the Raman peak positions are subject to shifts caused by the presence of water molecules  \cite{Song202013187, Prestipino20061337, Bae2022} which are sensitive to local heating induced by the laser illumination. The increased variability in the spectral range between 1100-1800 $rel. cm^{-1}$ could also be due to local heating. The band at 2920 $rel. cm^{-1}$, which was assigned to surface contaminants or adsorbants, also shows higher variability in the present case. Note, that this signal was absent in some reference crystals taken from the same batch.

In contrary, the bands between 400 and 1100 $rel. cm^{-1}$ which are either related to the vibration of the benzene rings or to the $OCO$ vibrations show little spectral variability. The peak position of the Raman band at 1007 $rel. cm^{-1}$, which represents the in-phase breathing vibration of the aromatic rings, shows the lowest variability in our analysis. Due to its high robustness and stability, we have chosen it as a spectral normalization reference to allow for analysis of crystal-to-crystal variations. The spectral peak positions and line widths, along with their respective standard deviations, are provided in TABLE \ref{tab:2}.

For quantifying crystal-to-crystal variability occurring in the same sample batch, we have acquired Raman spectra at the center position of 100 spatially isolated single HKUST-1 crystals. The averaged spectrum, see Figure \ref{fig:intra}b, shows noticeable variability in the low-frequency regime below 400 $rel. cm^{-1}$, as well as in the spectral ranges between 1100-1800 $rel. cm^{-1}$ and 2800-3100 $rel. cm^{-1}$. This is in agreement with the spectral variability we have observed on the single-crystal level. 

As a key results of our study, the peak positions of Raman bands in the averaged spectra acquired (a) at 100 different positions of the same, single crystal and, (b) at the center positions of 100 different, single crystals, are consistent. However, spectral standard deviations are larger in case (a) than in case (b). This demonstrates that heterogeneity can be high within the same crystal, underscoring the relevance of spatially resolved investigations.

In table \ref{tab:2}, we summarize the spectral peak positions and line widths, along with their respective standard deviations. The data can serve as a reference for lab characterization of HKUST-1 samples.

\begin{table*}[t]
\centering
\resizebox{0.5\textwidth}{!}{%
\begin{tabular}{ccccc}
\hline
\textbf{\begin{tabular}[c]{@{}c@{}}Peak Center\\ ($Rel. cm^{-1}$)\end{tabular}} & \textbf{\begin{tabular}[c]{@{}c@{}}$\Delta$ Center\\ ($Rel. cm^{-1}$)\end{tabular}} & \textbf{\begin{tabular}[c]{@{}c@{}}FWHM\\ ($Rel. cm^{-1}$)\end{tabular}} & \textbf{\begin{tabular}[c]{@{}c@{}}$\Delta$ FWHM\\ ($Rel. cm^{-1}$)\end{tabular}} & \textbf{\begin{tabular}[c]{@{}c@{}}Rel . Height\\ (Arb. units)\end{tabular}} \\ \hline
95.0 & 0.6 & 13.6 & 2.4 & 0.2 \\
111.8 & 1.2 & 23.2 & 7.6 & 0.2 \\
132.5 & 3.1 & 33.8 & 10.5 & 0.2 \\
178.5 & 2.9 & 43.0 & 11.0 & 0.2 \\
204.1 & 8.4 & 52.6 & 12.1 & 0.1 \\
280.8 & 1.1 & 21.3 & 3.9 & 0.1 \\
456.0 & 7.9 & 30.7 & 15.2 & 0.0 \\
500.9 & 0.3 & 24.3 & 1.0 & 0.3 \\
745.6 & 0.3 & 11.4 & 1.0 & 0.3 \\
814.0 & 1.6 & 22.2 & 4.2 & 0.1 \\
829.5 & 0.1 & 8.9 & 0.5 & 0.8 \\
1006.9 & 0.2 & 9.0 & 0.5 & 1.0 \\
1219.0 & 0.2 & 15.4 & 0.5 & 0.1 \\
1275.6 & 2.2 & 21.5 & 6.7 & 0.0 \\
1386.0 & 2.3 & 13.4 & 7.8 & 0.1 \\
1447.8 & 1.9 & 31.9 & 2.8 & 0.2 \\
1463.3 & 0.6 & 17.0 & 2.2 & 0.3 \\
1547.0 & 0.7 & 26.2 & 2.1 & 0.2 \\
1610.9 & 0.4 & 25.5 & 1.4 & 0.2 \\
2915.0 & 1.2 & 142.9 & 8.3 & 0.0 \\
3089.6 & 0.3 & 37.3 & 1.3 & 0.1 \\ \hline
\end{tabular}%
}
\caption{Experimental Raman spectral peak positions and line widths, respectively, of HKUST-1. The spectral variability is provided as standard deviation. The data represents 100 single-shot Raman spectra taken at the center positions of 100 different, single crystals taken from the same sample batch.)}
\label{tab:2}
\end{table*}

\subsection*{Principal Component Analysis for Identifying Single-Crystal Image Features}

After having established that Raman imaging can be used to map surface heterogeneity in single HKUST-crystals, see Figure \ref{fig:raman2d}, we now explore if multivariate data analysis can be used to reveal spatial features that are not visible in the mode-selective Raman images.

\begin{figure}[t]
    \centering
    \includegraphics[width=\linewidth]{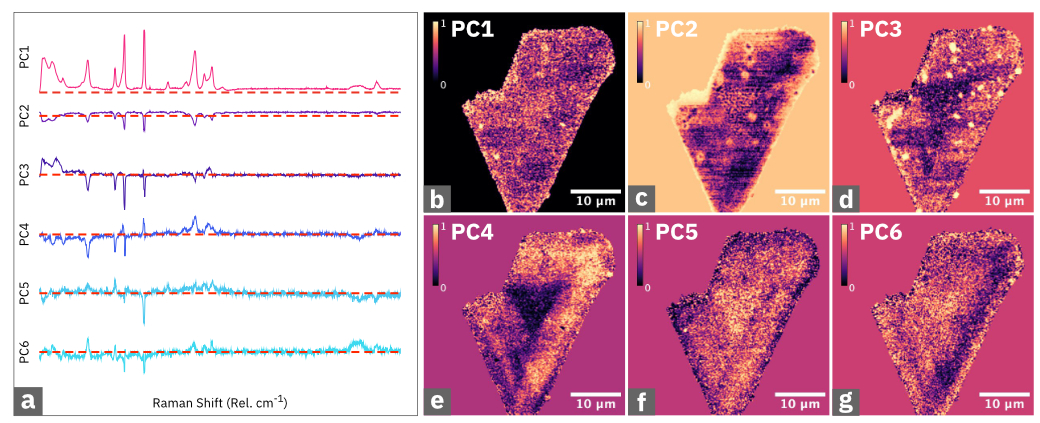}
    \caption{\textbf{Principal component analysis of HKUST-1 single-crystal Raman data.} (a) Spectral distribution of the six Principal Components (PC1-PC6). The dashed red line represents PC=0. (b - g) Images with spatial distribution of Principal Components 1-6 for the same HKUST-1 crystal.}
    \label{fig:pca}
\end{figure}

Principal component analysis (PCA) allows the transformation of higher-dimensional spectroscopic data into a lower-dimensional space and captures features within a select number of principal components \cite{pearson1901liii, hotelling1933analysis}. In the present case, the Raman image data shown in Figure \ref{fig:raman2d} was normalized and projected onto 6 principal components whose spectral distributions are plotted in Figure \ref{fig:pca}a. The spectral distribution of the first principal component, PC1, resembles the original HKUST-1 spectrum while PC2 is mainly a negative image of PC1. PC3 to PC6 have both positive an negative spectral contributions.

Component-selective Raman images were created by computing a 6-dimensional principal component space pixel-wise from the Raman spectral data. The dimensionality transformation operation yields a 6-channel data-cube where each channel represents one principal component, represented in Figure \ref{fig:pca}b-g. The images created for principle components 1 and 2, respectively, mainly exhibit contrast variations at the position of defects that are consistent in the mode-selective Raman images of Figure \ref{fig:raman2d}. However, the images for principle components 3-6 reveal a triangular area, having a side length of about 10 micron, which is located at the center of the crystal. This area was not readily visible in any of the mode-selective Raman images of Figure \ref{fig:raman2d}, although it is discernible in the normalized maps shown in Figure \ref{fig:sup1}.

In the images representing principle components 3 and 4, respectively, the triangular area exhibits negative image contrast, while the image contrast is positive in case of principle components 5 and 6. By looking at the spectral distributions of the principal components in Figure \ref{fig:pca}a, this could indicate a relative increase of spectral weight in the range between 500-750 $rel. cm^{-1}$  and a decrease of spectral weight in the region between 800-1000  $rel. cm^{-1}$. 


\section*{Summary And Conclusions}

In summary, we have performed an analysis of mode-selective Raman scattering in single HKUST-1 crystals. Based on ab-initio simulations, we have validated the principle Raman features of HKUST-1 and identified one Raman feature not previously reported elsewhere. We have microscopically resolved surface heterogeneities of spatially isolated, single HKUST-1 crystals that are caused by local defects and surface contamination. 

By analyzing Raman spectra acquired with the same, single crystal as well as with a set of different crystals, we establish a representative, single-crystal HKUST-1 Raman spectrum. The spectral peak positions and line widths, along with their standard deviations, can serve as a reference for monitoring MOF quality in the lab. The comparison of both sets of spectra reveals that spectral variability within the same crystal can be higher than across a set of different crystals taken from the same batch.

Finally, we have explored how principal component analysis can be applied for revealing features that cannot be straightforwardly identified in mode-selective Raman imaging. The analysis of the chemical origin of these features requires further research.

To allow for reuse and validation of our results, we have made the experimental data and the simulation code available in public repositories.

\section*{Acknowledgements}

F.L.O acknowledges financial support from CAPES (Project 001), CNPq, and FAPERJ. The authors acknowledge the use of computational infrastructure at Núcleo Avançado de Computação de Alto Desempenho (NACAD) of COPPE/UFRJ. 

\section*{Supporting Information}

The experimental and simulation data can be found in the following repository:
\begin{description}
    \item \textbf{Zeonodo:} \url{https://zenodo.org/uploads/14165785}
\end{description}

The code for the simulation of the Raman spectra of MOFs is provided in:
\begin{description}
    \item \textbf{GitHub:} \url{https://github.com/neumannrf/electronic-structure-experiment}
\end{description}

Further information on the Raman simulation methods and the mode selective Raman maps can be found in the supporting information.

\renewcommand\refname{References}
\begin{footnotesize}
\bibliographystyle{unsrt.bst} 
\textnormal{\bibliography{localbibliography.bib}}
\end{footnotesize}
\newpage

\end{doublespacing}

\title{\textit{Supporting Information:} Mode-selective Raman imaging of metal-organic frameworks reveals surface heterogeneities of single HKUST-1 crystals}
\maketitle

\begin{doublespacing}
\begin{linenumbers}
\setcounter{equation}{0}
\setcounter{figure}{0}
\setcounter{table}{0}
\setcounter{page}{1}
\resetlinenumber

\renewcommand{\thefigure}{S\arabic{figure}}
\section{Raman Simulations}

The vibrational modes were calculated under the harmonic approximation, where the potential energy surface ($U$) obtained by density functional theory is expanded in a Taylor series

\begin{equation}
   U(q_k) = U_0 + \sum_k \left ( \frac{\partial U}{\partial \vec{u}_k}\right )\vec{u}_k + \frac{1}{2} \sum_{j,k} \vec{u}_j \left ( \frac{\partial^2 U}{\partial \vec{u}_j \partial \vec{u}_k}\right ) \vec{u}_k + ...
\end{equation}

where $U_0$ is the ground state energy of the stable geometry, $\vec{u}_k$ is the cartesian coordinate displacement of the $k_{th}$ atom, $\frac{\partial U}{\partial \vec{u}_k}$ is the gradient of the electronic energy along this direction, and $H_{j,k} = \frac{\partial^2 U}{\partial \vec{u}_j \partial \vec{u}_k}$ is the second derivative (or Hessian matrix) along these directions. When the stable geometry correspond to a minimum on the potential surface the gradient terms vanish. 

Phonons were computed by diagonalizing the dynamical matrix obtained from the variation of atomic forces due to finite atomic displacements of 0.005 \AA{}. The diagonalization process was carried out using the phonopy package.\cite{togo2023first, togo2023implementation}

For a Stokes process, the differential scattering cross section of the \textit{m-th} vibrational mode is given by\cite{long2002raman}

\begin{equation}
\frac{d\sigma}{d\Omega} \propto \frac{(\omega_0 - \omega_m)^4}{\omega_m} | \mathbf{e}_i \cdot \alpha^m \cdot \mathbf{e}_s |^2 (n_m + 1)
\end{equation}

where $\mathbf{e}_i$ and $\mathbf{e}_s$ are the unit polarization vectors of incident and scattered radiation, $\omega_0$ is the frequency of incident radiation, $\omega_m$ is the frequency of the phonon mode, $n_m$ is the Bose-Einstein distribution given by $n(\omega_m) = 1 / (1 - e^{-\hbar \omega_m / k_B T} )$ and $\alpha^m$ is the Raman tensor given by

\begin{equation}
   \alpha_{ij}^m = \sqrt{V} \sum_{\alpha, \beta} \frac{\partial \chi_{ij}}{\partial r_{\alpha, \beta}} \tau^m_{\alpha, \beta}
\end{equation}

where $V$ is the unit cell volume, $r_{\alpha, \beta}$ is the $\alpha$ coordinate of $\beta$-th atom in the unit cell, and $\tau^m_{\alpha, \beta}$ corresponds to the normal vector of the $m$-th vibrational mode on the $\beta$ atom. 

The linear dielectric susceptibility tensor, $\chi_{ij}$ is calculated for each distorted structure using the linear response method as implemented in CP2K. 

To simplify the dependence on the polarization vectors $\mathbf{e}_i$ and $\mathbf{e}_s$ the Raman invariants ($a$, $\delta$, and $\gamma$) can be defined based on the Placzek approximation, in terms of the Raman tensor $\alpha$, with $I^m_{tot} = |\mathbf{e}_i \cdot \alpha^m \cdot \mathbf{e}_s |^2$. The set of invariants consists of the mean polarizability $a$, the antisymetric anisotropy $\delta$ and the anisotropy $\gamma$, given by:

\begin{equation}
   a^2 = \frac{1}{9}\left[ |\alpha_{xx} + \alpha_{yy} + \alpha_{zz}|^2\right]
\end{equation}

\begin{equation}
   \delta^2 = \frac{3}{4}\left[ |\alpha_{xy} - \alpha_{yx}|^2 + |\alpha_{xz} - \alpha_{zx}|^2 + |\alpha_{yz} - \alpha_{zy}|^2\right]
\end{equation}

\begin{equation}
   \begin{aligned}
   \gamma^2 = \frac{1}{2}\left[ |\alpha_{xx} - \alpha_{yy}|^2 + |\alpha_{xx} - \alpha_{zz}|^2 + |\alpha_{yy} - \alpha_{zz}|^2\right] \\ + \frac{3}{4}\left[ |\alpha_{xy} + \alpha_{yx}|^2 + |\alpha_{zx} + \alpha_{xz}|^2 + |\alpha_{yz} + \alpha_{zy}|^2\right]
   \end{aligned}
\end{equation}

Thus, the Raman intensity with parallel ($I_{\parallel}^m$) and perpendicular ($I_{\perp}^m$) polarization for the $m$-\textit{th} mode can be calculated as

\begin{equation}
    I_{\parallel}^m = \frac{45 a^2 + 4 \gamma^2}{45}
\end{equation}

\begin{equation}
    I_{\perp}^m = \frac{3 \gamma^2 + 5 \delta^2}{45}
\end{equation}

and thus

\begin{equation}
    I_{tot}^m = I_{\parallel}^m + I_{\perp}^m
\end{equation}

To generate the simulated spectra the Raman intensities were convoluted with a Lorentzian curve of width $\gamma$ = 5.0 cm\textsuperscript{-1} given by

\begin{equation}
f(\omega) = \frac{1}{\pi} \left [\frac{\gamma}{(\omega - \omega^n)^2 + \gamma^2} \right ] \cdot I^n
\end{equation}

where $\omega^n$ is the position of the $n$-th Raman peak, $I^n$ is the intensity of the $n$-th Raman peak and $\gamma$ is the width of the peak.

\pagebreak
\section{Mode-selective Raman images of single HKUST-1 crystal}

\begin{figure}[h!]
    \centering
    \includegraphics[width=0.9\linewidth]{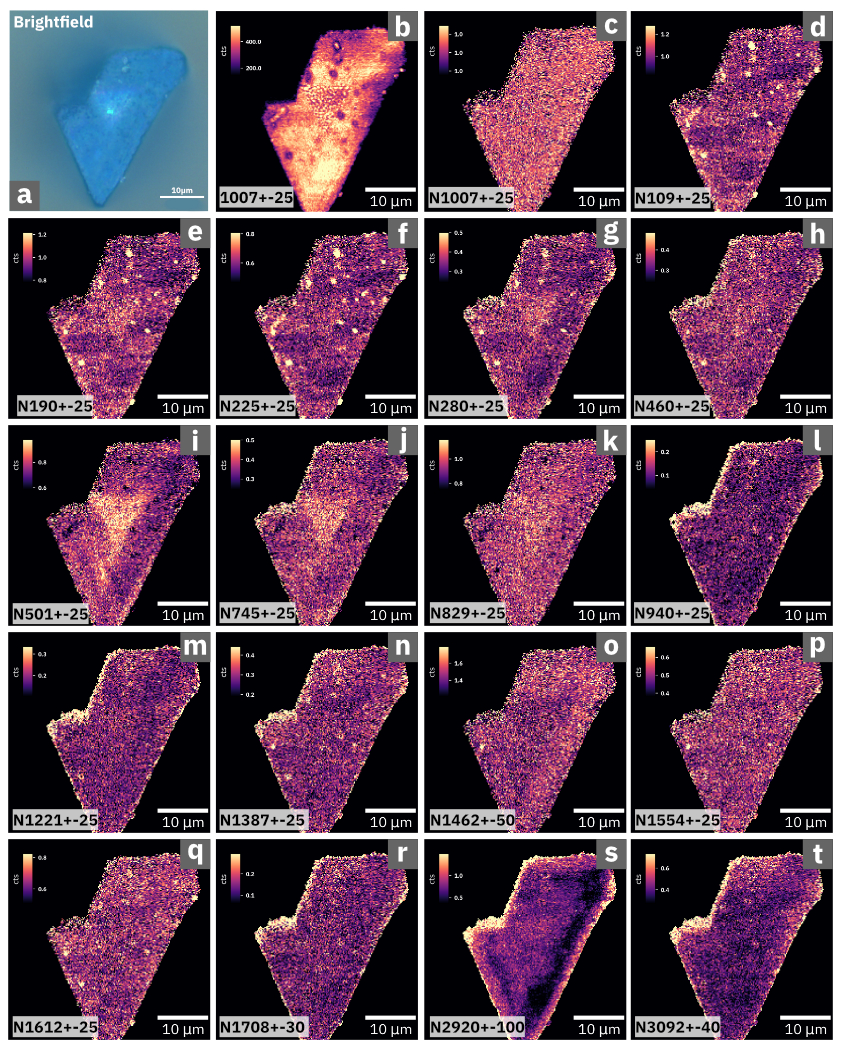}
    \caption{\textbf{Normalized Mode-selective Raman imaging of a single, spatially isolated HKUST-1 crystal}(a) Brightfield microscopy image of a single HKUST-1 crystal. (b) 1007 $\pm$ 25 rel. $cm^{-1}$ reference band. (c-t) Confocal Raman images of the same HKUST-1 crystal representing the normalized integrated intensities of select Raman bands in relation to the reference band}
    \label{fig:sup1}
\end{figure}

\renewcommand\refname{References}
\begin{footnotesize}
\end{footnotesize}
\newpage

\end{linenumbers}
\end{doublespacing}
\end{document}